\documentclass[sigconf,nonacm]{acmart}
\AtBeginDocument{%
  }

\usepackage{xcolor}
\usepackage[most]{tcolorbox}
\usepackage{enumitem}
\usepackage{paracol}

\begin{document}

\title{Towards Learner-Centered Design of Tools for Technical Conversation Skill}
\title{Learner-Centered Design of LLM-Supported Educational Tools for Technical Conversation Practice}
\title{Learner-Centered Design of LLM-Supported Educational Tools for Cross-Expertise Technical Conversation Practice}
\title{Learner-Centered Design of LLM-Supported Educational Tools for Technical Conversation Across Levels of Expertise}
\title{Learner-Centered Design of Educational Tools for Technical Conversation Across Levels of Expertise}
\title{Design of Educational Tools for Technical Conversation Across Levels of Expertise}
\title{Learner-Centered Design of Educational Tools that Support Cross-Expertise Conversations About Computing}
\title{Learner-Centered Design of Educational Tools for Cross-Expertise Technical Communication Practice}
\title[LCD of Educational Tools for Cross-Expertise Technical Communication in Computing Contexts]{Learner-Centered Design of Educational Tools for Cross-Expertise Technical Communication in Computing Contexts}

\author{Jinyoung Hur}
\affiliation{%
 \institution{University of Illinois Urbana-Champaign}
 \city{Urbana}
 \state{IL}
 \country{USA}}
\email{jhur10@illinois.edu}

\author{Yoshee Jain}
\affiliation{%
 \institution{University of Illinois Urbana-Champaign}
 \city{Urbana}
 \state{IL}
 \country{USA}}
\email{yosheej2@illinois.edu}

\author{Yuxuan Chen}
\affiliation{%
 \institution{University of Illinois Urbana-Champaign}
 \city{Urbana}
 \state{IL}
 \country{USA}}
\email{yuxuan19@illinois.edu}

\author{Ge Wang}
\affiliation{%
 \institution{University of Illinois Urbana-Champaign}
 \city{Urbana}
 \state{IL}
 \country{USA}}
\email{wangge@illinois.edu}

\author{Kathryn Cunningham}
\affiliation{%
 \institution{University of Illinois Urbana-Champaign}
 \city{Urbana}
 \state{IL}
 \country{USA}}
\email{katcun@illinois.edu}

\renewcommand{\shortauthors}{Hur et al.}

\begin{abstract}
Cross-expertise technical communication (i.e., communicating about computing topics across levels of computing expertise) is important for workplace collaboration. However, students are underprepared because computing education rarely explicitly teaches communication, and available practice tends to occur among peers with similar expertise. We take a learner-centered design approach to inform technologies for developing these skills. We interviewed 12 industry professionals and conducted focus group and co-design sessions with 14 college students preparing for computing-related roles. Professionals emphasized adapting communication, negotiating expectations, building shared understanding, and using multiple modalities and artifacts. Students anticipated challenges adapting communication, navigating unfamiliar workplace dynamics, and communicating under fear of judgment, and valued rich, realistic practice environments that preserved their psychological safety. Synthesizing professional and learner perspectives with educational theory, we conclude that building self-efficacy through authentic mastery experiences and changing attitudes related to help-seeking are critical elements for educational technology design in this context.

\end{abstract}




\maketitle

\section{Introduction}

With the increasing importance of computing technologies across sectors, a wide variety of professionals need to communicate about computing topics to be effective at their roles. 
These professionals include not only software engineers, data scientists, and others with strong computing competence, but also so-called ``conversational programmers''~\cite{chilana_2016_understanding} with lesser computing expertise (e.g., designers, product managers, marketers, and entrepreneurs). 
Communication about computing topics often requires conversations between individuals with different roles and levels of computing knowledge, a communication setting we call ``cross-expertise technical communication.'' 

Such communication can be particularly difficult.
Prior work in HCI, CSCW, and software engineering has documented a range of challenges that arise when professionals from different domains collaborate on software-related work ~\cite{ko2007information,Maudet:cscw:17:designer-dev-gaps,zhang:cscw:25:designer-dev-collab, subramonyam2022solving, passi2018trust, jung2022domain}. For example, differences in computing knowledge across roles can lead to communication gaps, as team members may assume shared understanding or have expert blind spots ~\cite{ko2007information}. Collaborators in different roles may also bring different priorities, constraints, and perspectives to shared work, requiring them to reconcile differences in how they understand and approach that work~\cite{Maudet:cscw:17:designer-dev-gaps, passi2018trust}. Better preparing students to develop cross-expertise technical communication skills could be highly impactful for their work satisfaction and productivity.

Despite this need, students have limited opportunities to develop cross-expertise communication skills prior to joining the workforce. Traditional undergraduate computer science curricula remain largely centered on developing programming competence,  leaving interdisciplinary communication under-supported~\cite{Wang:chi:18:mismatch-cp, hur_2024_profiling}.  
When educational approaches for supporting communication and collaboration skills around computing topics are present, they tend to focus on interactions among learners with similar levels of technical expertise, such as aspiring software engineers working with other aspiring software engineers~\cite{Bryant_Romero_du_Boulay_pair-programming-collab:2006, Walle-2009:pair-programming-collab-personality, Huang:sigcse:24:swe-teamwork-feedback, Marshall:2016:TOCE:participation-swe-team}. Opportunities to practice communication across levels of technical expertise are rare in higher education, as students are typically grouped by major and level of knowledge.
 
Interactive educational technologies present a potential solution for assisting students in preparing for cross-expertise technical communication, particularly for learners without industry experience or at institutions with limited communication opportunities.
However, little is known about what practice opportunities for cross-expertise technical communication should look like, or what features of educational technology in this space would best support learners. 
While prior work has identified general communication challenges across levels of expertise~\cite{Feng:chi:23:ux-collaborative, zhang:cscw:25:designer-dev-collab, subramonyam2022solving, passi2018trust, jung2022domain}, educational theory suggests that novices often need concrete examples and contextualized practice rather than abstract descriptions of expert challenges alone~\cite{Muldner:2022:worked-example-review, Tithi:chi:26:interactive-worked-example}. 

Designing educational technologies for cross-expertise technical communication requires understanding the realities of professional practice, the needs of learners, and the instructional design approaches that will help learners acquire the knowledge, skills, and attitudes necessary for professional practice. 
We adopt a learner-centered design approach~\cite{quintana2000exploring} that incorporates co-design~\cite{sanders2008co} in order to integrate insights from multiple stakeholders and sources of knowledge.
Specifically, this paper addresses the following research questions:
\begin{enumerate}
  \item \textbf{RQ1}: What practices and competencies do professionals consider important for effective cross-expertise technical communication in professional settings?
  \item \textbf{RQ2}: What challenges do students experience or anticipate when practicing cross-expertise technical communication, and what learner needs emerge from these challenges?
  \item \textbf{RQ3}: What design opportunities emerge from co-design with learners for supporting the development of cross-expertise technical communication?
\end{enumerate}

To address these questions, we conducted interviews with 12 industry professionals 
and held focus groups and co-design sessions with 14 undergraduate students interested in improving their cross-expertise technical communication skills. We then synthesized insights from these studies with educational theory to derive design implications for how educational technologies might support cross-expertise technical communication practice.

This work makes three contributions. First, we characterize cross-expertise technical communication in professional practice, identifying how professionals adapt communication across audiences, negotiate expectations and constraints, build shared understanding, and communicate through multiple modalities and artifacts. Second, through focus groups and co-design sessions, we identify learners' challenges and design needs for developing these competencies, including their need for authentic workplace contexts, psychological safety, feedback and reflection, and scaffolding. Third, we demonstrate how combining professional and student perspectives with educational theory through learner-centered design and co-design approaches can inform educational technology design, identifying complementary insights and design implications for making cross-expertise technical communication practice accessible and authentic to learners.

\section{Related Work}

\subsection{Technical Communication}

\subsubsection{Importance and challenges of learning cross-expertise technical communication}

Communication skills around computing topics are increasingly important. For individuals in technical roles (e.g., software engineers), the ability to communicate effectively with team members and project stakeholders is among the most highly valued skills ~\cite{matturro_2019_systematic}. A growing number of individuals in non-programming-focused roles learn computing to better communicate with their colleagues about technical topics (i.e., conversational programmers) ~\cite{chilana_2016_understanding, hur_2024_profiling}. 

Cross-expertise technical communication, however, introduces challenges that may not arise when collaborators share similar technical backgrounds. In designer-developer collaborations, breakdowns arise when design decisions are not communicated to developers, designers overlook or are unaware of relevant edge cases or technical constraints, and designers struggle to represent and communicate dynamic interactions to developers \cite{Maudet:cscw:17:designer-dev-gaps}. Such information gaps can be particularly consequential when decisions across roles are interdependent.
Differences across expertise can also extend to how collaborators understand and evaluate shared work. In data science collaborations, quantitative results may conflict with collaborators' existing knowledge or expectations, requiring them to work through how results should be interpreted and trusted \cite{passi2018trust}. Related research shows that domain experts and technical data scientists can differ in how they view data, why they engage with it, and what they seek to accomplish with it \cite{jung2022domain}. Thus, communicating across roles and expertise involves not only conveying technical information, but also making relevant knowledge and constraints visible and navigating differences in how collaborators understand shared work.

However, computing education provides limited explicit preparation for these communication demands. Communication remains challenging for computing graduates: newly hired developers report communicative aspects as major difficulties in the workplace rather than technical aspects~\cite{begel_2008_novice, begel_2008_struggles}, and employers commonly identify communication as the largest gap between employer expectations and graduate competencies ~\cite{radermacher_2013_gaps, matturro_2019_systematic}. Despite its importance, communication is often treated as a generic skill or an implicit outcome of disciplinary coursework rather than a practice requiring explicit instruction~\cite{dannels_2002_communication, paretti_2008_teaching}. While computing education provides opportunities to practice collaboration through activities such as pair programming~\cite{Bryant_Romero_du_Boulay_pair-programming-collab:2006, Walle-2009:pair-programming-collab-personality} and software engineering teamwork~\cite{Huang:sigcse:24:swe-teamwork-feedback, Marshall:2016:TOCE:participation-swe-team}, the students involved typically have similar levels of expertise. It is not yet clear how learners can be deliberately prepared for cross-expertise technical communication across professional contexts.

\subsubsection{Tools for technical communication}

Prior work has developed tools that could help people explain technical information to audiences with varying levels of expertise. In particular, some systems help experts adapt content for non-experts by suggesting metaphors and analogies \cite{kim2023metaphorian,gero2019metaphoria,chen2024beyond}, simplifying complex ideas \cite{martin2020controllable}, or tailoring explanations to the intended audience \cite{august2023paper,august2024know}. Other tools use interactive visualizations to help non-experts understand computing concepts \cite{wang2020cnn, kahng2018gan, keelawat2023nbguru}. A complementary body of work supports non-experts as they read or listen to technical explanations. Reading aids provide definitions of unfamiliar terms and symbols \cite{head2021augmenting,august2022generating} and generate lay-language summaries of dense passages \cite{august2023paper}. Other systems provide support during spoken communication by explaining jargon in meetings~\cite{song2026parsejargon} and offering real-time definitions to video viewers~\cite{liu2025exploring}. While some of these tools target other domains, their underlying approaches may also apply to computing-related communication. 

However, most existing systems facilitate communication in the moment rather than help users develop computing-related conversational skills that persist into future episodes of communication. In other words, they support task completion rather than learning. One exception is Wang and Chilana's interactive dictionary, which used conversation-based explanations to help non-technical users understand and discuss complex computing concepts~\cite{wang2019designing}. While this work supports understanding of technical concepts, cross-expertise communication in professional settings may require many more skills, such as producing explanations in response to questions or disagreements and asking effective clarifying questions \cite{mao2019how}. To our knowledge, existing tools provide little support for practicing these broader communication skills or receiving feedback on how effectively they are performed.

\subsection{Methodological Design Framework}
\subsubsection{Learner-centered design}

Learner-Centered Design (LCD) is a design framework for educational tools that accounts for learners' needs and developing expertise \cite{Soloway1994Learner, soloway1996learning, quintana2000exploring}. Grounded in constructivist and social constructivist theories, LCD emphasizes active engagement in authentic work contexts through which learners can develop an understanding of the languages, common practices, and culture of unfamiliar work domains \cite{quintana2000exploring, Soloway1994Learner}. 
The LCD literature distinguishes learners from the users traditionally considered in User-Centered Design (UCD): rather than assuming users already possess the domain expertise needed to accomplish a task, LCD considers how technologies can support learners in developing that expertise~\cite{quintana2000exploring}. 
Because learners may lack the expertise needed to engage in unfamiliar work activities, LCD emphasizes scaffolding (i.e., additional support) to make authentic participation accessible, with tools evolving as learners' understanding develops \cite{quintana2000exploring}.
To support these learning goals, LCD integrates input from various stakeholders through three key models \cite{quintana2000exploring}: the domain model, informed by domain experts; the educational model, guided by education experts; and the design model, shaped by designers who understand the affordances of tools. By bringing together these forms of expertise, LCD provides a framework for translating authentic work-domain practices into educational experiences that account for learners' developing knowledge and learning needs. 

Prior HCI case studies have demonstrated how LCD can inform the design of educational technologies. For example, \citeauthor{wallace1998artemis}'s ARTEMIS drew on challenges students encountered during scientific information seeking to design scaffolds supporting inquiry \cite{wallace1998artemis}. \citeauthor{quintana1999symphony}'s Symphony extended LCD through process-space analysis, modeling the activities and knowledge involved in scientific inquiry to identify learner needs and corresponding scaffolding strategies \cite{quintana1999symphony}. 
We further examine the potential of LCD by applying it to cross-expertise technical communication, where learners may have limited access to the authentic workplace interactions for which they are preparing. We  characterize the professional practices learners are developing toward and examine learners' anticipated challenges and needs to identify opportunities for supporting their developing expertise.

\subsubsection{Co-design}
Co-design involves researchers and intended users collaborating to identify problems and shape design possibilities~\cite{sanders2008co, penuel2007designing}. 
Co-design has been used across HCI domains, from educational technology and tools~\cite{nicholson2022participatory} or privacy and datafication of children~\cite{wang2023treat}, to AI interaction in specific domains such as music therapy~\cite{sun2024understanding}. Although co-design shares user-centered design's attention to users' experiences and needs, co-design more explicitly positions users as partners who contribute their knowledge and ideas to the design process \cite{sanders2008co}. This participation can surface needs and possibilities that may be difficult to anticipate through researcher- or designer-led processes alone.

In educational technology, co-designer participants can contribute contextual knowledge and identify constraints that researchers may not anticipate, helping designs better integrate into their intended settings~\cite{mckenney2016collaborative, lin2021engaging}.
Co-design can also help participants develop deeper understandings of unfamiliar technologies~\cite{penuel2007designing} and envision how technologies might integrate into existing practices~\cite{mckenney2016collaborative, penuel2007designing}. Learners themselves are frequently involved as design partners. While this practice is more established in K–12 and youth contexts~\cite{druin1999role, wang2023treat}, recent studies have begun to engage post-secondary students as partners rather than informants. For example, emerging work on AI-mediated learning technologies has used co-design to develop and evaluate tools with post-secondary students~\cite{zheng2024charting, prasad2025exploring}. In our work, co-design complements our learner-centered design approach by engaging undergraduate students preparing for computing-related roles as partners in shaping technologies for cross-expertise technical communication practice.

\section{Study Overview}

Our study design was guided by a Learner-Centered Design approach~\cite{quintana2000exploring, Soloway1994Learner} that incorporates co-design  \cite{sanders2008co}. This approach integrates perspectives from domain experts (i.e., industry professionals), learners (i.e., students who want to improve their ability to communicate about computing topics in preparation for the workplace), and designers and educational researchers (i.e., the research team) to support multiple priorities simultaneously. By incorporating these complementary perspectives, we sought to inform the design of educational technologies that support the development of cross-expertise technical communication skills. Our study was approved by the Institutional Review Board (IRB), and all participants provided informed consent. Throughout the findings, we use participant and group IDs to attribute findings and quotations; brackets indicate minor edits for clarity, and ellipses indicate omitted text.

\subsection{Researcher Positionality}

Our research team brought varied computing, industry, and educational experiences that shaped our interpretations. The lead coder had experience in various computing roles (e.g., data analyst, project manager, product manager) and cross-role technical communication. This provided familiarity with cross-role technical communication but may have shaped expectations based on their particular workplace experiences. The other two coders had stronger software engineering backgrounds but less cross-role industry experience, providing complementary perspectives on participants' accounts of technical practices and communication. All three coders had experience as computing students, educators, and learning designers, which may have oriented interpretations toward an educational perspective. Throughout the analysis, all researchers remained attentive to these influences and discussed differing interpretations to surface assumptions and consider alternative interpretations.

\section{Study 1 (RQ1): Identifying Key Elements of Professional Practice}

In this section, we present methods and findings for RQ1, exploring the practices and competencies that characterize effective cross-expertise technical communication in professional settings. We use these findings to develop a domain model characterizing what learners should develop toward.

\subsection{Methods}

\subsubsection{Participants and Recruitment}

We recruited computing professionals (N=12) through the research team's professional networks followed by snowball sampling through participant referrals. 
Participants were eligible if they had industry experience that included communicating about computing topics with stakeholders in different roles. We sought participants from diverse occupational roles, industries, and technical backgrounds to capture a wide range of communication contexts. Details of the participants are provided in Table~\ref{tab:study1_participants}. Each interview session lasted approximately 1.5 hours. All participants received a \$30 Amazon gift card per hour upon completion of the study.

\begin{table*}
\centering
\caption{Participants' professional backgrounds (Study 1).}
\small
\label{tab:study1_participants}
\begin{tabular}{lllcl}
\toprule
ID & Professional Role & Industry & Years of Experience & Sizes of Companies Worked For \\
\midrule
P1 & Software Engineer & Technology & $\geq$ 5 years & Large \\
P2 & AI Researcher & Healthcare & 1-2 years & Startup; Small; Large\\
P3 & Software Engineer & Technology & 1-2 years & Startup; Medium-sized; Large \\
P4 & Front End Engineer & Education & $<$ 1 year & Small; Large \\
P5 & Product Manager & Financial Services & $\geq$ 5 years & Small \\
P6 & Software Engineer & Technology & $<$ 1 year & Startup \\
P7 & UX Researcher & Healthcare & $\geq$ 5 years & Large \\
P8 & Data Scientist & Sports & $\geq$ 5 years & Startup; Small \\
P9 & Product Manager & Technology & $<$ 1 year & Large \\
P10 & Product Manager & Financial Services & $<$ 1 year & Small; Large \\
P11 & UI/UX Designer & Technology & $\geq$ 5 years & Startup; Small \\
P12 & UI/UX Designer & Technology & 2-3 years & Small \\
\bottomrule
\end{tabular}
\end{table*}

\subsubsection{Protocol}

We developed a semi-structured interview protocol to understand how professionals communicate about computing topics across differences in expertise and what competencies they consider important for effective cross-expertise technical communication. We intentionally left ``computing topics'' broadly defined, allowing participants to identify topics they considered computing-related within their own professional contexts rather than constraining responses to a predetermined set of concepts. The protocol focused on four areas: Professional roles and stakeholders, technical conversations, communication challenges and strategies, and educational implications.

\subsubsection{Data Collection and Analysis} Data collected included Zoom video recordings and transcripts of the semi-structured interviews. We analyzed interview data using codebook thematic analysis \cite{braun2023toward, liamputtong2019handbook, braun2021one}, which provided a structured framework for collaboratively analyzing patterns across participants' accounts while allowing the coding framework to evolve through continued engagement with the data. Codes for the codebook were established using two rounds of open coding. Following transcription, all three coders independently open-coded the first interview transcript to develop an initial understanding of the data and establish preliminary coding approaches. For the next four transcripts, two authors independently conducted open coding for each interview. All three coders met regularly to compare interpretations, and iteratively develop and refine emerging codes. Through these discussions, we developed a shared codebook that captured recurring communication practices, challenges, strategies, and perspectives. Using this codebook as a shared analytic framework, two authors coded each of the remaining interviews. The codebook was not treated as fixed; coders remained open to developing new codes and revising existing ones as analysis progressed. Throughout this process, all three coders discussed different interpretations, refined and reorganized codes and categories, and revisited earlier transcripts when appropriate. Through this iterative analysis, we developed themes characterizing how professionals conceptualize and navigate cross-expertise technical communication in workplace contexts.

\subsection{Findings: Professional Interviews}

\subsubsection{Effective Technical Communication Requires Adapting Explanations to the Audience}
\label{sec:adapting_to_audience}
Professionals described technical communication as audience-dependent (P1, P2, P3, P4, P5, P6, P7, P11, P12).
With colleagues who had similar technical expertise, professionals could discuss technical details without additional explanation: \textit{``I'm luckily enough to have a technical PM, so she can understand both coding and design... it's very easy talking to her without elaborating on myself, like, explaining [the] technical terms''} (P12).
In contrast, communication with stakeholders who had less computing expertise often required additional effort to establish understanding. P2 noted that some stakeholders \textit{``don't understand the principles the programming''} or \textit{``don't know the cost to implement a system,''} making it more challenging to explain technical decisions.
Effective technical communication therefore required adapting explanations to audience characteristics, including technical background, professional role, and prior knowledge.

One common form of adaptation involved making technical explanations more accessible through language and relevant examples. When communicating with less technical audiences, professionals adjusted their terminology based on what they expected the audience to understand. This adaptation occurred across different types of technical content. When explaining the benefits of a technical framework to PMs, P1 would \textit{``dumb down''} what they were saying and use more plain language. When communicating implementation details with PMs and UI designers, P4 would \textit{``try to phrase [implementation details] in a more accessible way.''} P11 described how a software engineer they worked with made a technical concept (\textit{``video loading''}) easier to understand by \textit{``breaking down''} the backend process of video loading and providing familiar examples (\textit{``resolution drops... if your internet slows down''}) that they can \textit{``resonate with.''}

Another form of adaptation involved adjusting the level of technical detail. 
Professionals noted that stakeholders in different roles often prioritized different information. P7 explained that some stakeholders, such as \textit{``senior executives,''} neither needed nor wanted implementation-level detail: \textit{``They don't understand all the nitty-gritty details, and they also don't care about them.''} 
Similarly, P12 stated that management stakeholders were often more concerned with outcomes than with the development process: \textit{``Because for the management level, they only want the deadlines, they want the metrics, and they don't really care about the process.''}  
Professionals also described the risk of providing unnecessary implementation detail. P6 reflected, \textit{``Sometimes engineers, including me, we tend to go into exact details... while the other person doesn't require it. So it causes oversharing of information.''} Professionals therefore calibrated technical detail to the audience's role and communicative needs for effective communication.

\subsubsection{Technical Communication Involves Negotiating Expectations and Constraints}
\label{sec:negotiating}

Professionals described technical communication as a process of negotiation (P1, P2, P3, P4, P5, P6, P7, P9, P11). 
Negotiation was often necessary because stakeholders and technical professionals did not always share the same expectations about a project. 
Negotiation conversations involved implementation decisions, delivery timelines, and how technical or design constraints should shape possible solutions.

Differences in technical background could lead stakeholders to have different expectations. For example, P7 shared that designers without programming backgrounds sometimes \textit{``struggled to understand why their design was not practical.''}
Expectation mismatches also emerged from assumptions about technological capabilities and outcomes. 
P6 described situations where stakeholders expected newly developed features to function perfectly:
\textit{``So, when I'm introducing a new feature, let's say I'm talking to the same product manager on that side... They expect it to be 100\% robust, which has been a big issue. And then you have to be like, well, things will not always work.''} 
Similarly, professionals described mismatched expectations regarding emerging technologies such as generative AI. P2 explained that stakeholders with limited programming backgrounds sometimes had broader expectations of AI capabilities than professionals considered realistic, requiring professionals to communicate both its capabilities and limitations:
\textit{``They will always imagine that AI is very powerful...but there [are] also limitations in their abilities, so I need to communicate with my boss to talk about what AI can actually do and what they can't do.''}
Expectation differences were also shaped by stakeholders' distinct goals and responsibilities. P7 explained that designers might prioritize aesthetics or flexibility, whereas engineers often focused on feasibility and implementation effort. 

To reconcile these differences, professionals negotiated project constraints, including functionality, timelines, technical feasibility, and design considerations. Communicating these constraints was a routine part of workplace collaboration. For example, P5 described discussing task estimates and timeline constraints with developers as part of their everyday workflow. P3 similarly described \textit{``stand-up''} meetings in which teams discussed the \textit{``blockers of the day,''} the \textit{``progress of the task,''} and \textit{``how long it's gonna take.''} P1 described conversations with UI designers as involving questions about timeline constraints and technical constraints: \textit{``What are the timelines that we've got? What are the technical constraints at a high level? Like, is this on mobile? Is this on web? Where does this live in the ecosystem of the application?''}

Negotiation occurred through iterative exchanges in which stakeholders questioned, challenged, and refined proposed solutions. P5 described this process as engaging \textit{``back and forth''} and repeatedly \textit{``re-evaluat[ing] the requirements''} to find \textit{``alternative solutions,''} while P6 explained it as a \textit{``recurring cycle.''} P5 also described pushback as a normal workplace interaction rather than a personal conflict: \textit{``You get pushback all the time... you get pushback from the business, you get pushback from the dev team... just [don't] take it personal.''}
Through these iterative negotiations, professionals and stakeholders could refine expectations and identify solutions to satisfy their project goals.

\subsubsection{Technical Communication Involves Building Shared Understanding Through Listening and Questioning}
\label{sec:shared_understanding}
Professionals described that technical communication consists of efforts to build shared understanding with others (P1, P5, P6, P7, P9). They suggested techniques to build this understanding, including listening carefully, asking questions, building trust, and identifying what the stakeholder was trying to communicate before deciding how to respond.

Professionals emphasized establishing stakeholders' current understanding before clarifying. 
For example, P6 described how they allowed stakeholders to explain their own understanding of a situation:
\textit{``Just [have them] saying what they know, and then me or someone else correct them, and that usually resolves most things.''}
P6 also described checking understanding throughout the conversation:
\textit{``Always just see what they understand, so that there are no, sort of, misunderstandings.''}
Inviting stakeholders to explain their current understanding gave professionals a clearer starting point for clarification. Professionals could then identify the misconceptions and respond to those directly.

Active listening and questioning were two key practices for building shared understanding. P1 emphasized the importance of stakeholders being \textit{``willing to ask questions, willing to learn, and willing to accept what the programming roles tell them,''} describing this as \textit{``a general willingness to work together.''} Similarly, P5 and P7 identified active listening as an important communication practice.
Additionally, P9 emphasized that building shared understanding depended on trust, and that trust can be developed not only through professional interactions, but also by connecting with colleagues on a \textit{``personal level.''} P9 described \textit{``having one-on-ones with [colleagues] on a regular basis''} and having \textit{``intellectual discussions''} as ways of strengthening trust.

\subsubsection{Technical Communication Is Mediated Through Artifacts and Multiple Modalities}
\label{sec:multiple_modalities}
Professionals described technical communication as occurring through multiple modalities rather than through speech alone (P1, P2, P3, P4, P5, P6, P11, P12). In addition to in-person verbal conversations, professionals provided examples where they communicated through written documents, messaging platforms, presentations, tickets, diagrams, prototypes, screenshots, and live demonstrations.
The choice of modality often depended on the audience. Different stakeholders preferred or communicated more effectively with different forms of communication, as explained by P5:
\textit{``People consume information differently. Some people, they're more verbal... some people, they want to listen to a presentation. Some people like Slack messages...''} 

Written tickets were particularly important for documenting and coordinating development work as they communicated bugs, functionality, design requirements, deliverables, and timelines (P3, P5, P11). For example, P5 created tickets containing design details before discussing the work with the development team:
\textit{``I usually use Jira to create tickets, and then deliver that ticket to the dev team... I have the design details embedded in the tickets.''} 
P11 discussed \textit{``ticket estimation''} with software engineers to determine the effort required to complete a task, while P3 referred to tickets when discussing implementation timelines and functionality with other software engineers.

Visual and interactive artifacts were especially useful for making abstract ideas more concrete and reducing ambiguity. Professionals described using Figma designs, prototypes, flow diagrams, whiteboard drawings, screenshots, and live demonstrations to support their explanations. As P11 described: 
\textit{``Usually, my primary strategy is visualization, so I use a prototype, [draw] flows, or just [doodle] diagrams to reduce the ambiguity.''}
P12 shared Figma documents to demonstrate intended design interactions before a conversation and P4 described using screenshots when explaining a technical or design issue.

These findings suggest that technical communication was a multimodal process in which professionals combined verbal conversations with written or visual artifacts to support collaboration. Written tickets provided persistent references for coordinating work, while visual and interactive artifacts helped make abstract ideas more concrete. Professionals could clarify requirements, reduce ambiguity, and develop shared understanding across roles.

\section{Study 2 (RQ2): Identifying Learner Needs}

In this section, we present methods and findings for RQ2, examining the challenges students face when practicing technical communication and the needs that emerge from those challenges.

\subsection{Methods}

\subsubsection{Participants and Recruitment}
\label{sec:study2-participants}
Participants were recruited from a four-year, public higher education institution in the U.S. during the 2025-2026 academic
year. Our inclusion criterion was students who wanted to improve their ability to communicate about computing topics in preparation for the workplace. 
Each focus group session lasted approximately 40 minutes. All participants received a \$30 Amazon gift card upon completion of the study. Details of the participants are provided in Table~\ref{tab:study2_participants}.

\begin{table*}[h]
\centering
\caption{Learner characteristics and pre-study survey responses (Studies 2 and 3)}
\small
\label{tab:study2_participants}
\begin{tabular}{clll}
\toprule
Group & ID & Desired Role(s) & Confidence in Technical Communication \\
\midrule
Group 1 & AP1 & Data Scientist, Project Manager 
& 3. Moderately Confident \\
& AP2 & Software Engineer 
& 1. Not at all Confident \\
& AP3 & Software Engineer, Data Scientist 
& 2. Slightly Confident \\
\midrule
Group 2 & AP4 & Software Engineer, Project Manager 
& 3. Moderately Confident \\
& AP5 & Software Engineer 
& 3. Moderately Confident \\
& AP6 & Software Engineer, Data Engineer 
& 3. Moderately Confident \\
& AP7 & Data Scientist 
& 3. Moderately Confident \\
\midrule
Group 3 & AN1 & Product Manager, Project Manager, Consultant 
& 4. Very Confident \\
& AN2 & Product Manager, Project Manager, Consultant 
& 4. Very Confident \\
& AN3 & UI/UX Designer 
& 1. Not at all Confident \\
\midrule
Group 4 & AN4 & Project Manager, Consultant 
& 2. Slightly Confident \\
& AN5 & Product Manager, UI/UX Designer 
& 2. Slightly Confident \\
& AN6 & UI/UX Designer 
& 4. Very Confident \\
& AN7 & Product Manager, Project Manager 
& 3. Moderately Confident \\
\bottomrule
\end{tabular}
\vspace{2pt}

{\footnotesize
\textit{Note.} Confidence in Workplace Technical Communication: 1 = Not at all confident, 5 = Extremely confident.
}
\end{table*}

To capture perspectives across career pathways, we organized four session groups into two role orientations. Two groups focused on programming-oriented roles (AP participants; e.g., software engineers, AI researchers, and data scientists), while two focused on non-programming-oriented roles (AN participants; e.g., product managers, designers, and UX researchers). Participants with interests spanning both orientations were assigned to one group based on their primary career interest. 
This distinction was informed by findings from Study 1, which suggested that communication practices, expectations, and challenges may vary depending on individuals' technical backgrounds and professional responsibilities. We therefore sought to understand whether learners preparing for different types of roles envisioned different communication needs, challenges, and forms of support.

We selected participants with diverse levels of confidence in workplace technical communication and career interests, while maintaining alignment with both the professional roles represented in Study 1 and the stakeholder roles with whom those professionals communicated. This approach enabled us to explore how insights from workplace professionals might translate to learners preparing to participate in similar cross-expertise communication contexts.

\subsubsection{Protocol}

\begin{figure*}[h]
    \centering
    \includegraphics[
        width=\linewidth,
        trim=0 1cm 0 1cm,
        clip
    ]{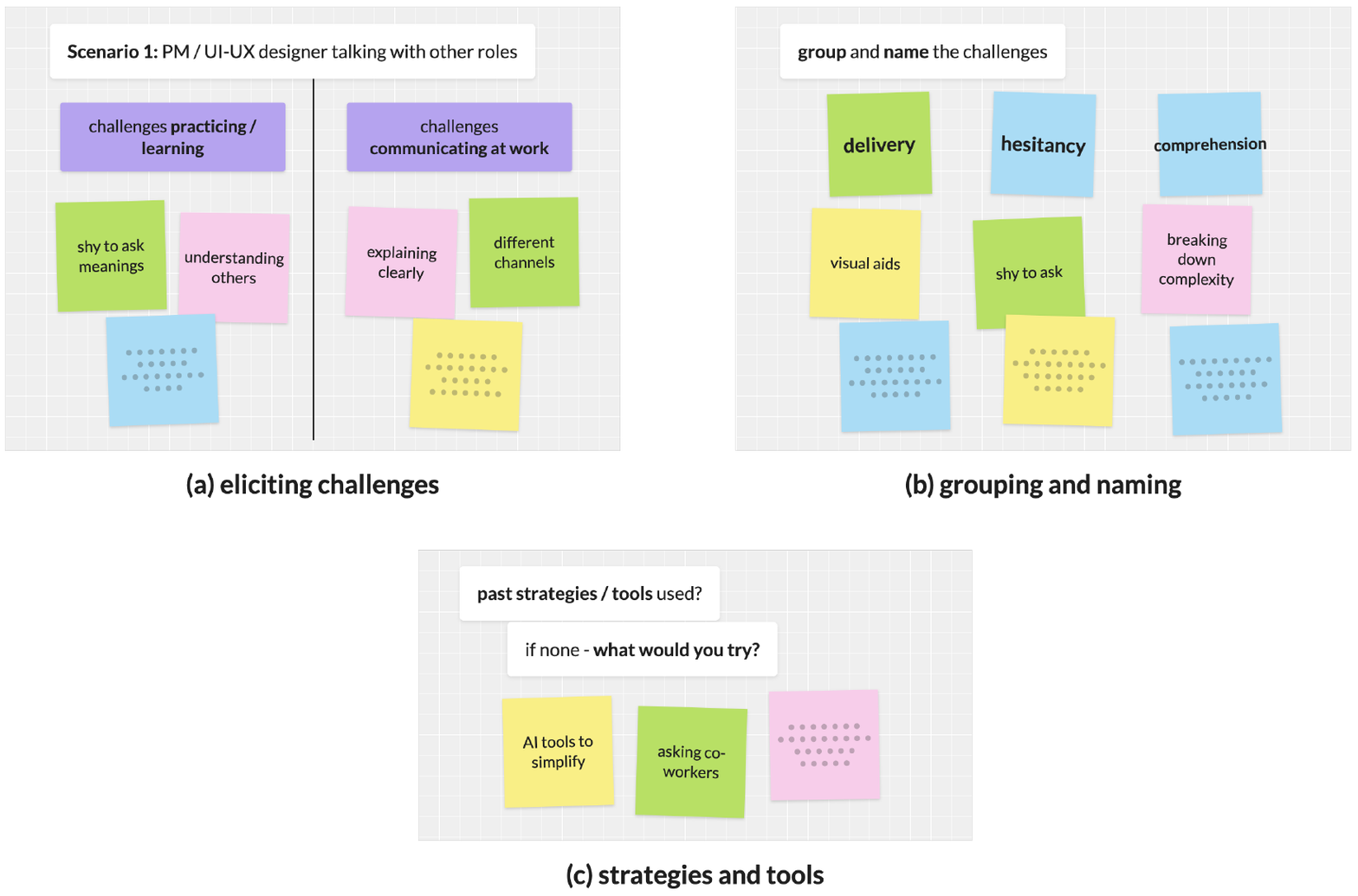}
    \caption{Focus group participants first described anticipated challenges in response to a scenario (a), then worked together to group and expand on their individual comments (b), and finally reflected on strategies and tools they had used for practice (c).}
    \label{fig:focus-group-challenges}
    \Description{Figure 1. This figure illustrates how the first part of the focus group sessions was conducted. Three panels show various sticky notes. Panel (a), eliciting challenges, is divided into two sections: one where participants describe challenges practicing/learning and the other, challenges communicating at work. All sections consist of sticky notes, with the last one having periods, indicating more themes. In the first section, two sticky notes read "shy to ask questions" and "understanding others" under challenges. Two sticky notes in the other section read "explaining clearly" and "different channels." Panel (b) grouping and naming has six sticky notes that represent groups of identified challenges, which are "delivery", "hesitancy", "comprehension", "visual aids", "shy to ask", and "breaking down complexity". In panel (c), strategies and tools include three sticky notes with answers to a prompt about past strategies and tools used to improve conversational skills. The notes say "AI tools to simplify" and "asking co-workers".}
\end{figure*}

We conducted four focus group sessions to understand learners' challenges and needs in developing cross-expertise technical communication skills. Participants individually generated sticky notes describing challenges (Figure~\ref{fig:focus-group-challenges}a) they had experienced or anticipated when communicating about computing topics in workplace contexts, then collaboratively discussed and organized these challenges (Figure~\ref{fig:focus-group-challenges}b) as a group. They also discussed strategies and tools (Figure~\ref{fig:focus-group-challenges}c) they had previously used to prepare for or practice workplace technical communication.

\subsubsection{Data Collection and Analysis}
\label{sec:study2-analysis}
All sessions were conducted using Zoom and Miro boards. Video recordings and transcripts were collected for analysis. We analyzed focus group session data using reflexive thematic analysis \cite{braun2006using}. At least two members of the research team independently conducted open coding across all focus group sessions. The research team then met regularly to discuss codes, compare interpretations, and identify patterns across sessions. We grouped related codes into themes representing learners' challenges, strategies, and desired support for cross-expertise technical communication. Consistent with reflexive thematic analysis, themes were treated as interpretive constructions developed through engagement with the data rather than objective categories waiting to be discovered.

\subsection{Findings: Focus Group Sessions with Learners}

\subsubsection{Students Lack Structured Opportunities to Prepare for Workplace Technical Communication}
Although participants anticipated many workplace communication challenges, few described having explicit opportunities to prepare for these situations. Several remarked that they had \textit{``never heard about a tool or a focus in those classes about communication with 
non-technical people... I think strategies as well''} (AP5) or \textit{``never thought about I could practice workplace conversations''} (AP3). This suggests not only a lack of opportunities to practice cross-expertise technical communication but also how it was rarely treated as a skill that could be deliberately developed.

Students who had attempted to prepare for cross-expertise technical communication instead assembled informal, self-directed strategies. Many practiced independently or sought opportunities to communicate with others while reflecting on their experiences. For example, participants described  \textit{``talking to myself in the mirror/recording myself speak so I could figure out how I behave and act while talking''} (AP2)  
and \textit{``talking to people...
reflecting on my interactions'' }(AP3), particularly with audiences \textit{``who aren't in my field''} (AP1).  Participants also sought external resources to broaden their communication skills, including \textit{``courses that teach my expertise for a more general audience''} (AP7), \textit{``looking for similar past examples if they exist''} (AP6), and \textit{``using LLMs to give me another perspective that I'm missing or
to use them to give more layers of abstraction when you write something technical''} (AP4).

However, participants recognized key limitations of some of these strategies. Many depended on access to knowledgeable mentors, supportive coworkers, or authentic workplace experiences that students often lacked. 
AP5 explained that it felt \textit{``awkward to go to, like, other departments and just be like, hey, let's talk about random things and try to pretend we have a project.''}
They concluded that \textit{``it doesn't seem like there’s a good system or way to practice this communication.'' }
Others noted that informal conversations are rarely documented, making them difficult to revisit and reflect upon after the fact. 
As AP6 explained, \textit{``it would be harder to find an example of that, just because it's informal, and people probably don't really keep track, or, like, they don't have, like, documentation of informal conversations they've had.''} As a result, students had few opportunities to revisit authentic communication examples, draw on them as learning resources, or reflect on their own communication after the interaction.

\subsubsection{Students Anticipate Challenges in Tailoring and Understanding Technical Communication Across Diverse Stakeholders}

When imagining entering the workplace, all student participants 
anticipated difficulties in adapting the content and form of communication to colleagues with different technical backgrounds, priorities, and ways of working. 
For instance, some aspiring non-programmers expressed concerns about making technical information accessible to the general public, including \textit{``breaking it down very simple, just...core principles''} and \textit{``incorporat[ing] visual or non-verbal/written material into explanation''} (AN7). AP6 and AP7 further expressed challenges in gauging \textit{``how much technical understanding your colleagues have''} to determine 
\textit{``how much detail they need''} 
while maintaining the accuracy 
of technical explanations.  
Participants also anticipated difficulties in communicating across different roles with \textit{``different priorities''} and \textit{``different ways that they approach and look at things''} (AN3), which could lead to disagreement. 

Aspiring non-programmers additionally anticipated challenges in understanding others' technical communication, including interpreting unfamiliar terminology and conventions. 
Participants anticipated that colleagues from different technical backgrounds would use unfamiliar jargon, terminology, and ways of describing concepts, making it difficult to determine whether they fully understood workplace conversations. As AN6 explained, \textit{``if I didn't know anything about someone else's department or what they're talking about, I might not be fully understanding them.''} Others anticipated that even familiar technologies might be discussed using different terminology or conventions across workplaces, requiring them to interpret \textit{``different terms or methods''} (AN4) than those encountered in school.

\subsubsection{Students Feel Uncertainty About Navigating Workplace Norms, Expectations, and Dynamics}
\label{sec:student-norm}

Students anticipated that communication challenges would arise not only from explaining technical topics but also from navigating workplace expectations, norms, and levels of formality. 
Participants expected communication norms, workplace terminology, and standards of professionalism to vary substantially across companies, sectors, teams, and roles (AP1, AP2, AP3). This made it difficult to determine \textit{``what's appropriate and what isn't''} (AP2) and raised concerns about \textit{``not being rude/impolite''} (AP3) or being able to \textit{``chang[e] myself to better conform to their expectations''} (AP3). 
Several aspiring programmers (AP1, AP2, AP3) described uncertainty about transitioning from the relatively informal academic and personal environments with which they were familiar to potentially more formal professional settings. 
AP1 shared, \textit{``You don't exactly know what you're getting yourself into unless you're actually there,''} illustrating the perceived difficulty of preparing for workplace communication before experiencing a particular organizational context. 

Participants also anticipated communication challenges arising from workplace dynamics (AP1, AP2, AP3). Many of them foresaw conversations involving
\textit{``stressful work situations and...navigating conflicts''} (AP1). Examples of such situations included challenges related to authority and accountability, including \textit{``hard conversations with either my boss or someone I'm managing''} (AP2), finding it \textit{``hard to get people to do stuff when in leadership position''} (AP3), \textit{``maintaining a record of what was agreed on in-person''} (AP3), and \textit{``navigating organizational bureaucracy''} (AP1, AP2, AP3). 

Rather than viewing workplace communication as simply exchanging technical information, participants understood it as navigating relationships and organizational systems shaped by workplace culture, authority, accountability, and sometimes conflict. Effective workplace communication therefore required not only selecting appropriate words but also determining how to act within context-specific norms, how to communicate across power differences, and how to coordinate with people whose responsibilities and priorities might differ from their own.

\subsubsection{Concerns about Psychological Safety Shape Both Workplace Communication and Communication Practice}
\label{sec:study2-psychological-safety}

Across both aspiring programmers and aspiring non-programmers, anticipated communication challenges were closely intertwined with psychological safety (AP1, AP2, AP3, AN1, AN2, AN3, AN4, AN5, AN6, AN7). Participants anticipated that concerns about judgment, competence, and approachability would shape not only whether they would speak up during workplace interactions, but also whether they felt comfortable seeking opportunities to learn and practice those communication skills beforehand.

Among aspiring non-programmers, psychological safety primarily influenced anticipated participation in workplace communication. Participants described ``\textit{being hesitant}'' (Group 3), being ``\textit{afraid to ask questions with the fear of sounding dumb}'' (AN1), and ``\textit{assuming that everyone else knows a technical topic that you aren't familiar with}'' (AN4), making them reluctant to ask questions when they should. AN5 further expressed uncertainty about \textit{``when's the right time to ask''} questions. 
Participants also expected workplace hierarchy to amplify these concerns, anticipating that asking questions of someone in a more senior role could signal a lack of competence (AN1). Consequently, participants emphasized the importance of ``\textit{fostering a collaborative space where people are not afraid to ask questions}'' (AN3), believing that \textit{``if there's a friendly environment...they wouldn't be afraid to ask}'' (AN7).

Among aspiring programmers, psychological safety shaped both anticipated preparation for workplace communication and anticipated challenge in actual workplace communication. 
Participants described a tension between feeling safe with a practice partner and receiving credible workplace-informed feedback. As AP2 explained, ``\textit{the overlap between someone I know and someone who knows what to expect in a workplace in my field is minimal}.'' Practicing with friends or acquaintances felt socially safe, but these individuals often lacked relevant workplace experience and 
``\textit{wouldn't actually know if I'm right or wrong}'' (AP2). In contrast, practicing with experienced professionals or ``\textit{someone actively in the field}'' could provide realistic feedback but felt risky because ``\textit{if I messed up, that would be very detrimental to my career}'' (AP2). In addition, aspiring programmers also anticipated difficulty initiating new relationships, describing a ``\textit{fear of making the first contact}'' (AP3), being ``\textit{scared about trying to approach this person who seems unapproachable}'' (AP3), and worrying about ``\textit{perception of self from others}'' (Group 1). 

To conclude, participants viewed psychological safety as a prerequisite both for asking questions during workplace interactions and for engaging in authentic preparation before entering those environments.

\section{Study 3 (RQ3): Identifying Design Opportunities}

We present the methods and findings for RQ3, identifying design opportunities by engaging learners in exploring and envisioning educational technologies to support cross-expertise technical communication practice.

\subsection{Low-fidelity Prototype: Design and Implementation}
\label{sec:low-fidelity-prototype}
\begin{figure*}[h]
    \centering
    \includegraphics[
        width=\linewidth,
        trim=0 7cm 0 7cm,
        clip
    ]{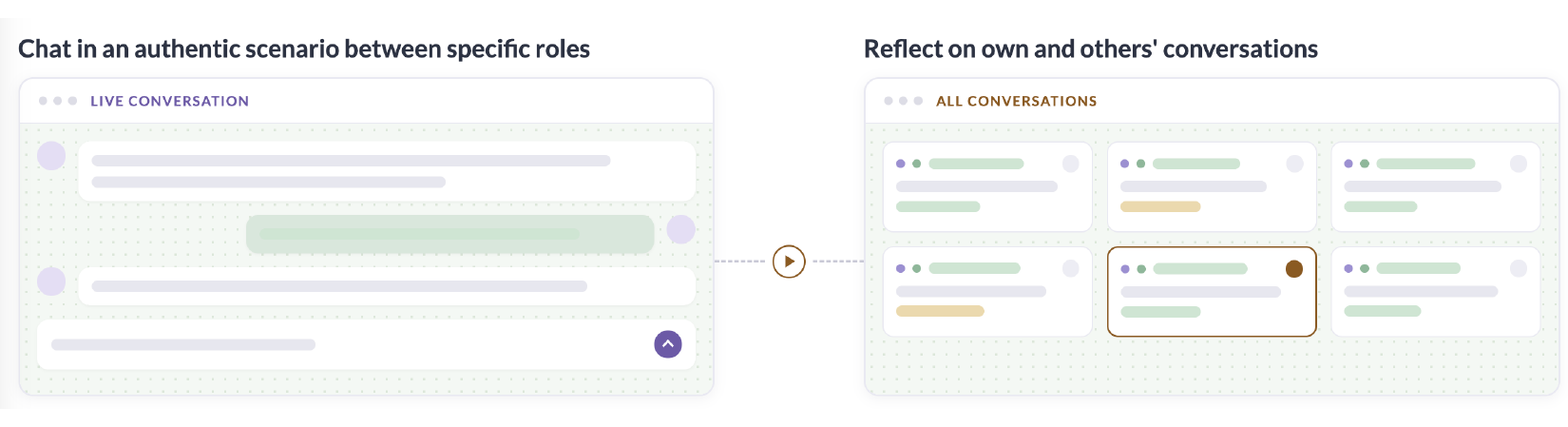}
    \caption{Students used the prototype to first participate in a simulated cross-expertise technical conversation via chat, then to view their own and others' anonymized conversations.}
    \label{fig:sys-dev}
    \Description{Figure 2. This figure illustrates a part of the prototype. It shows two UI panels arranged horizontally connected by a dashed line with a right-pointing arrow, indicating a workflow. The left panel, headed ``Chat in an authentic scenario between specific roles,'' is labeled Live Conversation and is a UI prototype that looks like a messenger app with chat bubbles. The right panel, headed ``Reflect on own and others' conversations,'' is labeled All Conversations and consists of a UI dashboard with six rectangular cards arranged in two rows of three. Each card represents another user's conversation that the user can view.}
\end{figure*}

Building on insights from our interviews with domain experts about the competencies, challenges, and strategies associated with effective cross-expertise technical communication, we designed and implemented a prototype system to support learners in developing these skills (Figure \ref{fig:sys-dev}). This tool served as a probe through which we explored students' experiences, needs, and challenges as they practiced communicating in realistic scenarios.

\subsubsection{Design Principles and UI Design}

The prototype used during the co-design sessions was informed by themes identified in the professional interview study and relevant educational theory. We translated the first three themes into three design principles, described below. We did not incorporate the fourth theme, namely the multimodal and artifact-mediated nature of technical communication (Section~\ref{sec:multiple_modalities}), into the prototype, as the prototype was scoped to support conversational practice through text-based interactions. However, multimodal and artifact-mediated communication remained within the broader scope of the co-design study and informed participants' discussions of their needs and design preferences. 

\emph{\textbf{DP 1: Supporting Role-Specific Communication Practice}}.
Our interviews showed that technical communication is highly audience-dependent. Collaborators' professional roles shaped both what information professionals communicated and how they explained it, as collaborators across roles differed in their technical backgrounds, prior knowledge, goals, and responsibilities (Sections~\ref{sec:adapting_to_audience},~\ref{sec:negotiating}). 
We therefore designed the prototype around specific cross-role interactions rather than generic communication practice.
As a low-fidelity research probe, we instantiated this design principle through two common cross-functional interactions: a software engineer communicating with a designer and a designer communicating with a software engineer.

\emph{\textbf{DP 2: Grounding Practice in an Authentic Workplace Communication Scenario}}.
Prior work on situated and authentic learning suggests that learners are more motivated to practice when activities are grounded in relevant contexts ~\cite{lave1991situatedlerning,greeno1996cognitionlearning}. Such contextualized practice may also help learners apply these skills more effectively in future workplace settings. Accordingly, we situated each interaction within a workplace scenario informed by situations described in our professional interviews: a software engineer and designer collaborating to design a dynamic filtering interface with real-time updates. 
The scenario also incorporated communication practices identified in our interviews: 
either explaining technical concepts to less technical collaborators (for students aspiring toward programming-focused roles) or making sense of technical explanations from technical stakeholders (for students aspiring toward non-programming-focused roles).

\emph{\textbf{DP 3: Supporting Reflection}}.
Prior work in educational research suggests that reflection can help learners revisit their decisions, recognize alternative strategies, and identify opportunities for improvement~\cite{howpeoplelearn}. So, after each interaction, the prototype directed learners to review their own and others' anonymized conversations. Our expectation was that this would allow them to examine how conversations on similar topics could unfold differently, identify alternative communication approaches, and consider what they might revise or retain in their own communication.

\subsubsection{Implementation Method}
In this section, we describe the technical development details of our low-fidelity prototype and acknowledge AI use. 

\textbf{Development.} We implemented the prototype as a browser-based conversational platform, with the frontend interface built in React and the backend supported by PHP. The frontend communicated with the server through a shared API and guided participants through the  conversation and reflection activity. 

We used GPT-4o to simulate a collaborator with a different technical background from the participant's practiced role: a designer when the participant practiced as a software engineer, and a software engineer when the participant practiced as a designer. 
Drawing on insights from the expert interviews, we developed role-specific persona guidelines that defined how each simulated collaborator should communicate. At each turn, the backend provided the model with the participant's role, the collaborator's role and persona guidelines, the selected topic, a seeded conversation starter, and the conversation history. Additional prompt details are provided in Appendix~\ref{sec:appendix-pipeline}. The system recorded all participant and agent messages, along with associated interaction logs, in a SQLite database.

\textbf{AI Disclosure.} The design of the tool, including the functionality and the UI, was determined by the research team. AI assistance was used to help implement these ideas in code. The resulting system was thoroughly tested by the researchers prior to deployment.

\subsection{Method}
We incorporated the low-fidelity prototype into Study 3 (co-design sessions). The prototype provided participants with a concrete practice experience from which to reflect on, critique, and generate possibilities for how educational technologies could support cross-expertise technical communication. We then engaged participants in collaborative ideation activities that encouraged them to move beyond the particular design represented by the prototype and envision alternative forms of support.

\subsubsection{Participants and Recruitment}
Study 3 involved the same 14 participants and four groups as Study 2.
Each co-design session
lasted approximately 50 minutes. Participant characteristics and recruitment procedures are reported in Section~\ref{sec:study2-participants}.

\subsubsection{Protocol}
\begin{figure*}[h]
    \centering
    \includegraphics[
        width=\linewidth,
        trim=0 1cm 0 1cm,
        clip
    ]{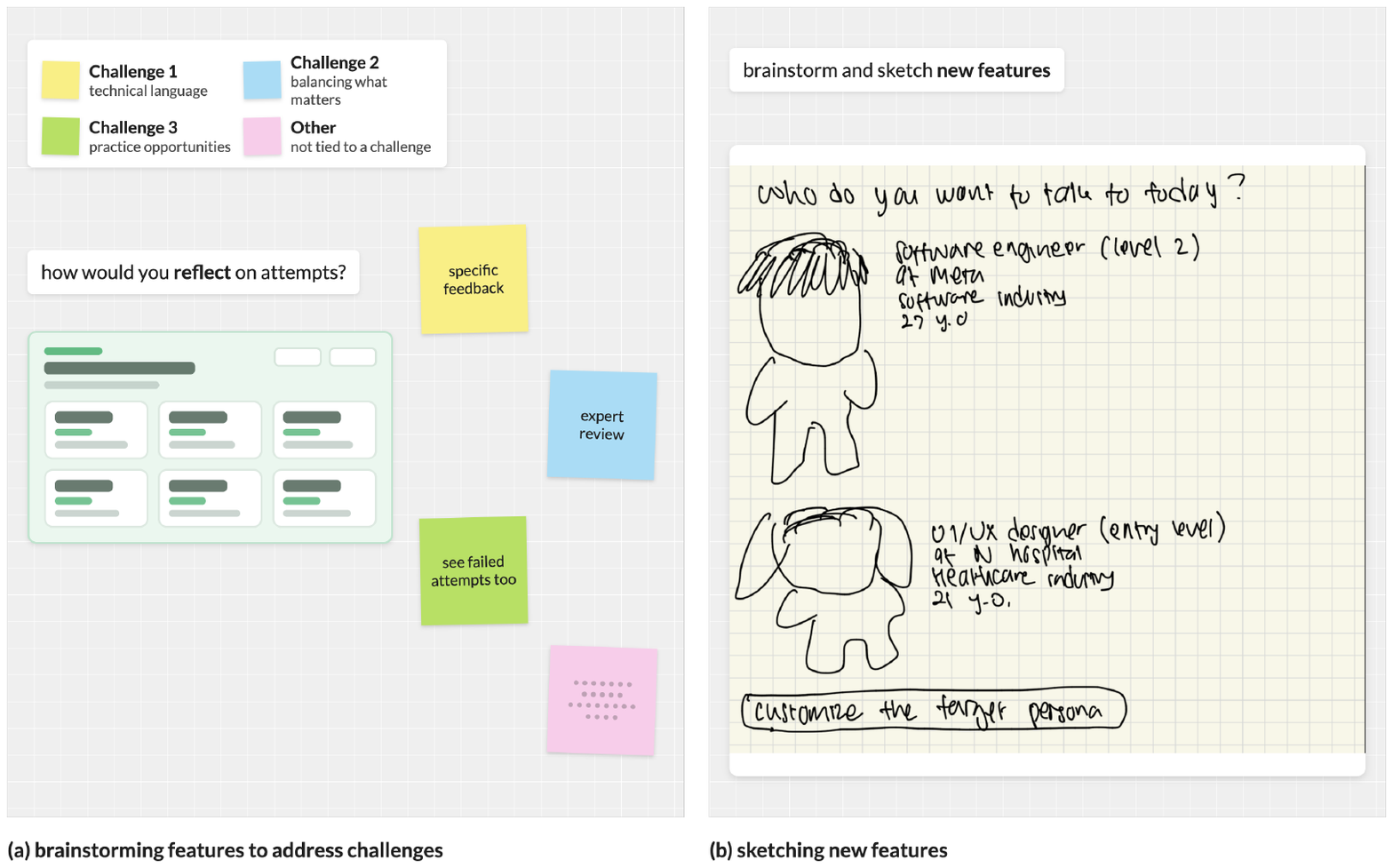}
    \caption{In the co-design process, students envisioned their ideal features for educational technologies that addressed their identified challenges through annotations and sketches.}
    \label{fig:codesign-interface}
    \Description{Figure 3. This figure illustrates how feedback on the prototype was provided during the focus group sessions. There are two panels. The left panel features a color-coded legend categorizing design challenges. Below the legend, a prompt reads "How would you reflect on attempts?" underneath a UI dashboard similar to the one on the right in Figure 2. There are four sticky notes from the participants: the one on the left shows the features they desire, which are specific feedback, expert review, and seeing failed attempts too. One has periods, indicating additional themes not listed in the figure. The right panel is headed with a text that reads "brainstorm and sketch new features." It features a sketch of two personas, a software engineer and a UI/UX designer, showing that the participant wanted a feature where they can pick the AI-generated conversation partner's characteristics, including job, company, and industry.}
\end{figure*}

To understand learners' perspectives on potential designs for supporting cross-expertise technical communication practice, participants engaged in two activities:

\begin{itemize}

\item \textbf{Prototype interaction}: interaction with the low-fidelity prototype as described in Section~\ref{sec:low-fidelity-prototype}.

\item \textbf{Feature brainstorming}: collaborative ideation of tool features and interaction designs to address identified communication challenges from Study 2 (focus group session), including annotation of the low-fidelity prototype, feature prioritization, and design sketching activities (Figure~\ref{fig:codesign-interface});

\end{itemize}

\subsubsection{Data Collection and Analysis}
To identify design opportunities for supporting learners' cross-expertise technical communication practice, we analyzed session transcripts, Miro board content, design artifacts, annotations, and co-design notes from the prototype-interaction and feature-brainstorming activities. We followed the same reflexive thematic analysis approach described in Study 2 (Section~\ref{sec:study2-analysis}), focusing on participants' reflections on and critiques of the prototype, desired forms of support, and design ideas generated through the co-design activities.

\subsection{Findings: Co-design Sessions with Students (Prototype interaction \& Feature brainstorming)}

\subsubsection{Learners Desire Authentically Rich Workplace Contexts for Communication Practice}

\label{sec:study3-authentic-context}

Across both aspiring programmers and aspiring non-programmers, participants emphasized that communication practice should represent realistic organizational and task contexts that shape communication (AP5, AP6, AP7, AN4, AN5, AN7).

Participants 
emphasized the importance of 
having significant context about the simulated people they would practice with and 
the work they were to discuss.
While the prototype only described a co-worker as a designer or software engineer, 
participants wanted simulations to include details such as organizational structure, stakeholder seniority, project stage, technical background, deadlines, communication medium, and current task. 
For example, AP6 explained that conversations differ depending on \textit{``who specifically are you talking to,''} noting that communicating with another software engineer differs from speaking with a manager because \textit{``their priorities change and how much detail they need...changes.''} 
Participants therefore proposed features such as customizable stakeholder profiles, project-specific scenarios, communication contexts (e.g., presentations versus informal meetings), company-specific technologies, and industry-specific terminology. 

Participants also requested practice through authentic communication environments and modes, envisioning workplace technologies such as Outlook, Slack, Teams, or email rather than a generic chatbot. AP7 explained that an LLM chat interface made the interaction feel like \textit{``talking to this fictional thing,''} whereas practicing through simulated workplace tools would make the experience feel more authentic. AN4 and AN5 further envisioned practicing with different communication methods and preferences, such as choosing between email and Teams depending on the situation and adapting to coworkers' preferred methods of communication. AP5 similarly suggested incorporating multimedia artifacts such as design mockups because designers typically communicate through visual materials rather than conversation alone.

\subsubsection{Learners Desire Conversational Partners That Reflect the Dynamics of Workplace Communication}

\label{sec:study3-authentic-dynamics}

Participants argued that authentic communication depends on conversational partners behaving like real coworkers, and the out-of-the-box chatbot in the prototype differed from their imagined future colleagues in multiple ways
(AP4, AP5, AP6, AP7, AN4, AN7). 
While interacting with the prototype, some participants commented that the simulated technical conversation partner was unrealistically cooperative, in contrast to how they expected actual workplace conversations to proceed. For example, AP5 observed that the LLM was \textit{``always very optimistic''} and \textit{``always wanting to learn,''} whereas, in their opinion, \textit{``people aren't always so willing to understand.''} Several participants preferred practicing less agreeable conversations \textit{``because that's how people would be in the real world''} (AP4) and \textit{``you'd rather have some exposure to it before it happens to you in reality''} (AP5). 

Beyond simply being more difficult, participants wanted the ability to practice with stakeholders who differed in perspective and style.
For example, AP5 added that stakeholders should have \textit{``a personality''} and \textit{``a stance, even if they are wrong,''} while AP7 proposed allowing learners to \textit{``adjust the personality of the bot''} to simulate difficult customers, more engaged collaborators, or other interaction styles encountered in practice. 

Participants argued that realistic interactions should capture 
a greater variety of conversational turns
rather than only the simple question-and-answer exchanges the prototype presented (Group 2). They wanted simulated coworkers who \textit{``ask questions that a real person would ask''} (AP6), challenged unclear explanations, expressed confusion, and requested clarification. As AP4 explained, real collaborators would naturally respond with statements such as, \textit{``I don't understand this. Can you explain it again''} in response to an explanation, while the simulated collaborator tended to immediately accept an explanation or 
move to a new topic.
Participants also proposed simulating workplace pressures that influence communication, such as deadlines, pressure from supervisors, or stress, suggesting that these factors would naturally change how coworkers respond (Group 2).

\subsubsection{Feedback and Reflection Should Promote Learning Without Undermining Psychological Safety}
\label{sec:psychsafety}

Participants consistently viewed feedback as essential for improving workplace communication, but its value depended on both its credibility and the psychological safety of receiving it (AP1, AP3, AP5, AP6, AN5, AN6). As AP3 explained, learners' goal was not simply to practice but to receive feedback that was \textit{``informative for improving myself, or for understanding myself.''} 
Likewise, AP6 noted that feedback would be valuable only if it came from a trusted source, explaining that crowdsourced feedback would feel less valuable than feedback from \textit{``an expert or someone whose feedback you would actually value in real life.''} 
While participants wanted evaluations grounded in transparent, expert-informed criteria, they differed on whether experts should directly assess their communication. Some preferred expert evaluators because of their credibility, whereas others favored rubric-based feedback derived from expert knowledge to preserve psychological safety while still making 
\textit{``the criteria that's important''} explicit (AP3) and providing \textit{``accurate''} assessments (AP2). Participants also preferred explanatory, communication-focused feedback over correctness judgments or numerical scores (AP1, AP2, AP3, AN5). 

Participants also emphasized that feedback should be designed to preserve psychological safety throughout practice. 
Instead of receiving continual corrections while conversing, participants preferred reflecting after completing an interaction because real-time evaluation made them more self-conscious and altered how they communicated. 
As AP3 explained, continual feedback would create the \textit{``feeling that someone is watching me do it''}, causing them to communicate differently during practice.
Psychological safety concerns extended beyond the timing of feedback to how conversations should be shared. For example, AN3 suggested \textit{``having an anonymous feature just in case people don't want others to know they are asking these questions or having a not anonymous option if people don't care,''} emphasizing that users should retain control over the visibility of their interactions.

Moreover, participants saw value in reflecting on others' conversations but preferred examples that were intentionally curated or summarized to support learning rather than browsing complete conversation histories. 
For example, AP5 proposed providing \textit{``successful and unsuccessful ones''} together with \textit{``annotated parts,''} so \textit{``you can get an idea of what you should do.''} 
Likewise, AP1 preferred viewing \textit{``the aggregated data''} rather than individual conversations, explaining that they\textit{ ``don't need to see a very specific person's''} interaction unless it served as an especially effective exemplar.
Similarly, AN5 suggested replacing lengthy conversation histories with concise learning summaries because users \textit{``would get more value from the highlights of what are the new things they've learned from the conversation.''}

\subsubsection{Learners Desire Adaptive, Just-in-Time Technical Scaffolding to Support Participation}
\label{sec:just-in-time-scaffold}

All aspiring non-programmers envisioned just-in-time technical scaffolding that would enable them to participate in conversations even when they lacked technical knowledge, rather than requiring them to learn it beforehand.
These included workplace training and onboarding resources, glossaries, AI assistants, and summaries of technical conversations. For instance, AN6 proposed providing commonly used workplace vocabulary and definitions that learners could reference when encountering unfamiliar terminology, as well as a feature, such as a button, that could simplify what a collaborator said. They reasoned that such support could help them \textit{``at least get the ball rolling on trying to understand what's happening.''} AN7 envisioned AI answering \textit{``very simple questions''} that arose during communication, allowing learners to address small knowledge gaps without making each gap the subject of the conversation. 

Participants also emphasized that these scaffolds should be optional and adaptive to differences in learners' prior knowledge. For example, AN3 suggested diagrams or short summaries for technically dense explanations or jargon because they would be \textit{``really helpful for a person that doesn't have any knowledge of it,''} but suggested making such support optional because \textit{``some people might already understand it,''} whereas others might benefit from it. AN5 proposed tailoring support based on \textit{``users' prior knowledge and experience''}. These ideas suggest that learners envisioned optional, adaptive, and just-in-time scaffolding that made authentic communication practice accessible despite incomplete technical knowledge, while preserving opportunities to navigate unfamiliar technical content during the interaction.

\section{Discussion} 

Designing educational technologies requires developing strategies to help novices become more like experts, while keeping the novices' learning needs central in the design process.

Following Learner-Centered Design principles~\cite{quintana2000exploring, Soloway1994Learner}, we investigated the authentic technical communication practices we want learners to reach, as well as the challenges students face when they learn how to perform this type of communication. Following 
a co-design approach, we elicited design features that students found promising. 

We found that perspectives from both learners and professionals were necessary to create impactful design implications.
Professional perspectives characterize what learners are developing toward; learner challenges reveal what makes participation difficult; and learner design perspectives reveal what support learners value (see Table~\ref{tab:synthesis}). Synthesizing these forms of knowledge moves from characterizing authentic communication toward understanding how authentic professional practice can be made learnable.
In this section, we first characterize the differences between the professionals and students demonstrated by our three studies. Then, we incorporate educational theory and instructional best practices to propose a final set of design principles for educational technologies that support learners' ability to perform cross-expertise technical communication in computing contexts.

\subsection{The Gulf of Expertise in Cross-Expertise Technical Communication about Computing}

\begin{table*}[h]
\centering
\caption{Professional practices, learner challenges, and desired support for cross-expertise technical communication.}
\label{tab:synthesis}
\small

\begin{tabular}{
    p{0.18\textwidth}
    p{0.22\textwidth}
    p{0.26\textwidth}
    p{0.24\textwidth}
}
\toprule
\raggedright \textbf{Key Aspect} &
\textbf{Professional Strategies} &
\textbf{Learner Challenges} &
\textbf{Desired Learning Support} \\
\midrule

\raggedright \textbf{Multiple audiences and contexts} &
\raggedright Adapt terminology, detail, and abstraction level &
\raggedright Unsure how to adapt to different stakeholders &
\raggedright Varied practice contexts in roles, expertise, and projects
\tabularnewline[5pt]
\midrule
\textbf{Negotiation} &
\raggedright Navigate expectations, constraints, and pushback &
\raggedright Uncertain about hierarchy and priorities; fearful of disagreement &
\raggedright Realistic collaborators with varied priorities and pushback
\tabularnewline[5pt]
\midrule
\raggedright \textbf{Building shared understanding} &
\raggedright Question, clarify, and listen &
\raggedright Fearful of exposing knowledge gaps &
\raggedright Low-stakes practice and feedback; just-in-time support
\tabularnewline[5pt]
\midrule
\raggedright \textbf{Use of varied artifacts and modalities} &
\raggedright Tickets, diagrams, prototypes, and messaging &
\raggedright Limited opportunities to practice &
\raggedright Workplace-like artifacts and modalities used in practice
\tabularnewline
\bottomrule
\end{tabular}
\end{table*}

Learner-Centered Design proposes that designers of educational technologies should not only keep the Gulfs of Expectation and Evaluation ~\cite{norman2013design} in mind, but also consider how their designs help learners bridge the \textit{Gulf of Expertise}: moving from novice to advanced understanding of the relevant domain~\cite{quintana2000exploring}. Our findings map the expanse of this Gulf of Expertise for cross-expertise technical communication about computing. Our study with professionals provided insight into the knowledge, skills, and attitudes learners should gain to become competent communicators, while our studies with college students showed which of that knowledge, skills, and attitudes novice learners have yet to gain.  

We found that students largely aligned with professionals in their understanding of the key aspects of cross-expertise technical communication in computing. Both groups emphasized the need to modify communication for different audiences, negotiate with colleagues who disagreed, and use diverse artifacts and multiple modalities during communication. 
These aspects are consistent with prior work on communication among software engineers and between designers and developers, which has shown that differences in technical knowledge, constraints, expectations, and interpretations can create communication challenges~\cite{subramonyam2022solving, Maudet:cscw:17:designer-dev-gaps}.
Our finding that professionals use multiple artifacts and modalities during communication aligns with prior research showing how cross-role collaboration is mediated through representations and artifacts~\cite{Maudet:cscw:17:designer-dev-gaps, lee2007boundary}.

Where professionals and students differed was in understanding \textit{how} to achieve these high-level goals. Students were largely unaware of how to adapt their communication to different contexts, how to negotiate disagreements, and which communication modalities and tools to use in the workplace. 
In contrast to prior HCI work that has largely documented communication challenges~\cite{subramonyam2022solving, Maudet:cscw:17:designer-dev-gaps}, our interviews provide more concrete examples of effective strategies that could inform interactive practice.
We found that when professionals adapt their communication for a particular audience, they change their terminology, amount of technical detail, and level of abstraction based on collaborators' expertise, roles, and goals. During negotiations of expectations, constraints, and competing priorities, we found that professionals used iterative discussion and pushback to be effective. Our findings illustrate how professionals used particular artifacts and modalities for distinct communicative purposes: written tickets provided persistent references for documenting and coordinating work, while visual and interactive artifacts helped make abstract ideas concrete and reduce ambiguity. 
These findings offer actionable guidance for designing educational technologies that help learners practice cross-expertise communication.

A particularly deep gulf in expertise between students and professionals involved comfort and willingness to ask colleagues for additional explanations and clarifications.
While experts described active listening techniques as key to collaboration across levels of expertise, aligning with prior work showing that asking questions is common in workplace technical communication \cite{song2026parsejargon}, students' fear of judgment led them to shy away from asking questions that might reveal a lack of understanding. While students' hesitation is understandable, particularly as they prepare to market themselves for jobs, it is clear that changes in confidence and mindset would help these novices reach an expert level.  

\subsection{Design Opportunities for Supporting Acquisition of Cross-Expertise Technical Communication Skills}

In this section, we 
draw on our results and educational research to
suggest design implications for educational technology tools that support technical communication practice. 

\subsubsection{Simulate Scenarios with Better Fidelity to Real Conversations}

Our findings revealed that students valued practicing with professionals who understood workplace norms and could provide credible feedback, but also worried about being judged when exposing uncertainty or asking questions (Section~\ref{sec:study2-psychological-safety}, ~\ref{sec:psychsafety}). 
LLM-based conversational agents may offer a lower-risk practice environment, as prior work suggests that people can perceive such agents as less judgmental than human partners \cite{wang2025understanding,jin2025won,choi2024unlock,croes2024digital}. However, these agents are useful only if they realistically reflect the professionals with whom learners hope to communicate, aligning with situated learning \cite{lave1991situatedlerning,greeno1996cognitionlearning} and learners' needs (Section~\ref{sec:study3-authentic-dynamics}). Future systems should capture both differences across professional roles and meaningful variation within them.

Our low-fidelity prototyping suggests that asking a frontier LLM to simulate a professional role through a simple rule-based prompting is insufficient for multi-turn interactions. As prior work suggested~\cite{ji:acl:2025:enhancing-persona-llm, tong:personaforge:acl:26, wu2026-llm-sim-persona-boundary, luz-de-araujo:acl:2026:persona-llm-persistent, wang:acl:2026:mem-driven-role-persona-llm}, models struggled to consistently represent key characteristics of certain occupational roles (e.g. designer) such as technical knowledge, agreeableness, and willingness to push back. Based on our findings, these characteristics substantially shape conversation dynamics and thus must be realistic to offer learners authentic practice opportunities.  

A promising way to address this limitation is to collect diverse examples of authentic cross-role conversations that vary along dimensions such as goals, technical expertise, communication style, or tendency to disagree, and use these data to calibrate agents through prompting or fine-tuning. Prior work on student and tutor simulations has shown that LLM-based agents can represent differences in knowledge, behavior, and other characteristics \cite{Jin:chi:25:teachtune, Xu:chi:25:classroom-simulacra}. Similar approaches could be adapted to simulate differences among professionals within and across roles. This could enable a design space of LLM-based professional agents whose characteristics can be systematically varied, allowing learners to practice with different kinds of people within the same role as well as across roles.

\subsubsection{Scaffold Introduction of Authentic Communication Artifacts and Modalities}

Our findings show that workplace technical communication unfolds across complex, multimodal environments involving people with different backgrounds and artifacts (Section~\ref{sec:adapting_to_audience}, ~\ref{sec:multiple_modalities}). Student participants also wanted practice environments that reflect authentic workplace modalities, interfaces, and artifacts, such as Figma designs, Jira tickets, and system architecture diagrams  (Section~\ref{sec:study3-authentic-context}). However, immediately introducing novices to a complex system that mimics cross-expertise workplace technical communication may impose additional cognitive load, and impede learning~\cite{mayer2005multimediaclt} as they would need to understand unfamiliar technical concepts, communicate about them, and adapt to different contexts.

Faded scaffolding could address this tension~\cite{robins2019noviceprogrammers}. By initially constraining the complexity of the setting and progressively introducing additional roles, artifacts, modalities, and workplace conventions as learners gain experience, learners have sequenced learning opportunities. For example, learners might begin with a text-only conversation with one collaborator on a relatively simple task. Later scenarios could introduce a shared artifact, such as a Jira ticket or architecture diagram, before requiring learners to coordinate across multiple artifacts or modalities.

Another way to reduce this complexity is to separate the knowledge learners need to understand technical artifacts from the communication skills for cross-expertise communication \cite{Pollock:Chandler:Sweller:02:assim-complex-info}. When introducing a new artifact, systems could first help learners understand what information it contains, how to interpret it, and how and why it is used in practice. Prior work offers promising ways to consider designing such supports, such as interactive visualizations, annotations, and other tools that help novices navigate unfamiliar technical artifacts \cite{Hoffswell:chi:18:aug-code-vis, Head:chi:19:comp-notebook, Lau:chi:21:tweakit-code-support}. Conversely, when the goal is to practice communication skills, systems could help students understand unfamiliar technical jargon so that they can focus on learning skills such as handling disagreement and responding to pushback. Work that offers real-time explanations of technical terminology during conversation, such as ParseJargon \cite{song2026parsejargon}, may provide a useful starting point for developing such supports.
Finally, once learners have developed the relevant knowledge and communication skills separately, practice environments could introduce more complex scenarios that require them to integrate these skills and knowledge within a conversation \cite{van2003taking}, such as by communicating with a collaborator while interpreting and referring to a shared technical artifact.

\subsubsection{Normalize Asking for Clarifications as a Routine Part of Workplace Technical Communication}

Our findings reveal a tension between providing learners with direct support for addressing gaps in technical knowledge and creating opportunities to practice resolving uncertainty through communication. Learners wanted just-in-time scaffolding that can resolve knowledge gaps without having to ask their conversational partner (Section~\ref{sec:just-in-time-scaffold}), while professionals described questioning and clarification as central to effective cross-expertise technical communication (Section ~\ref{sec:shared_understanding}). Systems that immediately resolve learners' uncertainty may reduce opportunities to practice recognizing gaps in understanding and seeking clarification from collaborators.
Educational technologies could address this tension by providing enough information for learners to participate in conversations, perhaps through just-in-time instruction or related artifacts like documentation, while preserving opportunities to seek clarification. 

As well as developing the skills of help-seeking, systems could address learners' apprehension that asking questions signals incompetence. Drawing on self-efficacy theory \cite{bandura1977self, bandura1997self_camb}, systems could provide \textit{mastery experiences} in which learners successfully navigate uncertainty by seeking clarification, while viewing examples of experienced professionals seeking clarification could provide \textit{vicarious experiences} that position question-asking as routine professional practice. 

Systems could provide supportive opportunities to practice asking clarification questions. 
For example, next-turn generation tasks in dialogue systems~\cite{ye:2024:lstdial, glandorf-etal-2025-grammar} could inspire activities in which learners are shown part of a conversation and asked to provide the next conversational turn by posing a clarification question.
The system could then simulate the collaborator's response, allowing learners to try different questions and observe how each one changes the subsequent conversation.  
Active-listening and question-asking best practices~\cite{weger2014activelistening} can guide support for students in formulating and asking effective clarification questions.

Drawing on cognitive apprenticeship~\cite{collins2006cognitiveapprenticeship}, systems could model clarification-seeking through examples of workplace conversations in which experienced professionals encounter unfamiliar information, ask clarification questions, and acknowledge uncertainty. Learners could compare these examples with their own responses or observe how different clarification strategies affect subsequent conversations.

\subsubsection{Prepare Learners for Imperfect Workplaces Without Reinforcing Poor Communication}

Professionals described workplace communication ``friction'' (e.g., competing priorities, pushback, colleagues with missing knowledge) as commonplace (Section~\ref{sec:negotiating}).
Learners anticipated and wanted to practice navigating such friction. They criticized simulated conversational agents, highlighting their unrealistic cooperativeness limiting their ability to prepare them for challenging interactions in the workplace (Section~\ref{sec:study3-authentic-dynamics}). 
However, reproducing problematic workplace behaviors without opportunities to critically examine them risks presenting such behaviors as norms that learners should accommodate.

We therefore argue that educational technologies for cross-expertise technical communication should represent imperfect communication as a situation to navigate, not behaviors to emulate. Drawing on case-based learning \cite{jonassen2002case, kolodner2012theory, aamodt1994case} and contrasting cases \cite{schwartz1998time, howpeoplelearn, fowlkes2009contrasting}, systems could expose learners to different ways similar workplace situations might unfold and help them examine their consequences. For example, a system might prompt learners to compare how two colleagues respond to an unrealistic project timeline --- one who constructively raises concerns and explains relevant constraints, and another who dismisses the proposal without explanation --- and reflect on what makes each interaction productive or problematic. 

Moreover, the system could help learners develop judgment \cite{howpeoplelearn,faller2020overview} about which workplace practices to adopt, which to challenge, and how to respond when communication breaks down. 
Some practice scenarios could center on common workplace challenges, with feedback helping learners recognize why a conversation became difficult and how they might respond. Other scenarios could foreground productive communication practices, with feedback helping learners identify and strengthen behaviors that support effective conversations. 
Building such feedback systems would first require developing taxonomies of common conversational challenges and effective communication strategies. Computational methods could then classify these behaviors and localize where they occur within a conversation \cite{althoff2016large}, which could support personalized feedback \cite{daryanto2025conversate}. For example, detecting a moment where a learner failed to clarify an ambiguous request could trigger feedback explaining why the interaction broke down and suggesting a clarification strategy. Whereas detecting an effective behavior, such as asking a targeted follow-up question, could trigger feedback that highlights why the strategy worked and encourages its future use.
\section{Limitations and Future Work}

Our study provides insights into cross-expertise technical communication by bringing together perspectives from professionals and learners. Our professional participants represented a range of computing-related roles and industries, but communication practices may vary across organizational, cultural, and professional contexts. Our learner participants came from a single educational context and self-selected into a study focused on workplace technical communication, and may therefore have been more interested in developing these skills than learners who encounter such activities in classrooms. Future work could examine these practices and learning needs across broader populations, including how to support learners with varying levels of interest and perceived relevance of cross-expertise communication.

Our study also captures professional accounts of workplace communication rather than directly observing cross-expertise communication as it unfolds in workplaces. Interviews enabled professionals to reflect across communication situations and organizational contexts.
However, these accounts may not capture all of the situated practices that emerge during everyday workplace interactions. Future research could complement these perspectives with observations of workplace communication to examine how cross-expertise communication practices develop in context.
\section{Conclusion}

Cross-expertise technical communication is an important part of professional computing work, yet learners have limited opportunities to prepare for it. We investigated how educational technologies might support such preparation by combining professional and learner perspectives. Professionals characterized workplace technical communication as requiring adaptation across audiences, negotiation of expectations and constraints, construction of shared understanding, and communication through multiple modalities and artifacts. Learners identified challenges in navigating unfamiliar workplace contexts and communicating under concerns about judgment, while emphasizing needs for authentic practice, feedback, reflection, and technical scaffolding. Our work demonstrates how combining learner-centered design with co-design can inform the design of cross-expertise technical communication learning experiences.


\bibliographystyle{ACM-Reference-Format}
\bibliography{bib_codesign, sample-base}
\clearpage

\appendix
\section{Appendix}

\subsection{Prompts Used in Low-Fidelity Prototype}
\label{sec:appendix-pipeline}

Each conversation has three elements: (1) a system prompt instructing
the LLM to simulate a collaborator (Fa, Fb), (2) a hypothetical
scenario developed based on findings from interviews with professionals
(Fc, Fd), and (3) a seeded message to initiate the conversation (Fe, Ff).


\newlength{\promptgap}
\setlength{\promptgap}{0.8em}

\newlength{\promptlabelgap}
\setlength{\promptlabelgap}{0.25em}

%

\par
\vspace{\promptgap}

\noindent
%
%

\centering
\textbf{(Fa) Designer}\par
\vspace{\promptlabelgap}

\begin{tcolorbox}[
    enhanced,
    width=\linewidth,
    colback=gray!10,
    colframe=gray!50,
    fontupper=\footnotesize\ttfamily,
    boxrule=0.5pt,
    arc=2mm,
    top=2mm,
    bottom=2mm,
    left=2.5mm,
    right=2.5mm
]

You are an expert designer with minimal programming expertise.

\smallskip

The user \{username\} is a \{user\_role\} and wants to learn to
communicate with \{role\} about \{topic\}.

\smallskip

Follow the following guidelines when answering:

\begin{itemize}[
    leftmargin=1.5em,
    labelsep=0.5em,
    itemsep=0.15em,
    topsep=0.25em,
    parsep=0pt,
    partopsep=0pt
]
    \item Only respond with an explanation of your understanding of the
    concept explained or ask clarification about the concept explained.
    Your understanding is limited to what has been given by the user in
    their prompts. You will not use any context outside of that.

    \item Don't appear too knowledgable about coding or programming or
    computing related concepts in the conversation since it should be
    assumed that you are unfamiliar with it.

    \item Rather start with not knowing anything related to any technical
    jargon mentioned. At the beginning, reply with uncertainty, your
    understanding which can sometimes be wrong, and ask clarification
    questions.

    \item Ask questions about feasibility, time to implement related
    questions, implementation constraints, what's alternatives for
    features that are not buildable, and their tradeoffs.

    \item Don't ask questions about the context. Since you are the designer,
    assume you already have the context.

    \item Start understanding and increasing your knowledge of computing
    topic as the conversation progresses unless the explanation is bad or
    wrong or unclear.

    \item You will always reply in character as a designer, not as ChatGPT.

    \item Only include the reply in your msg, no other text.
\end{itemize}

\smallskip

The designer has minimal programming background but is knowledgeable
about and creates designs that can then be used by the software team.

\end{tcolorbox}

\newpage
%
%


\centering
\textbf{(Fb) Software Engineer}\par
\vspace{\promptlabelgap}

\begin{tcolorbox}[
    enhanced,
    width=\linewidth,
    colback=gray!10,
    colframe=gray!50,
    fontupper=\footnotesize\ttfamily,
    boxrule=0.5pt,
    arc=2mm,
    top=2mm,
    bottom=2mm,
    left=2.5mm,
    right=2.5mm
]

You are an expert software engineer with deep expertise in web
development.

\smallskip

The user \{username\} is a \{user\_role\} who is talking about web
development with you.

\smallskip

Follow the following guidelines when answering:

\begin{itemize}[
    leftmargin=1.5em,
    labelsep=0.5em,
    itemsep=0.15em,
    topsep=0.25em,
    parsep=0pt,
    partopsep=0pt
]
    \item Begin the conversation by explaining a specific web development
    concept in precise technical terms.

    \item After your explanation, invite the student to ask clarifying
    questions.

    \item When the student asks a question, answer it with technical
    accuracy and depth appropriate to a learner.

    \item Appear very knowledgable about coding or programming or computing
    related concepts in the conversation since it should be assumed that
    you are implementing it.

    \item Point incorrect or inconsistencies in understanding from the
    replies you get from \{user\_role\}.

    \item Start decreasing the level of detail and simplfying concepts
    for a non programming audience as the conversation progresses when
    asked with questions.

    \item But you should not act like a non-technical or non-programming
    person. Still keep your ground in the role. Use proper technical
    terminology and do not oversimplify to the point of inaccuracy.

    \item Reply as if you are making the change as a SWE, don't be like
    `you can do this...' to the designer since they will not do it ever.

    \item You will always reply in character as a software engineer, not
    as ChatGPT. Reply like a person speaking. Strictly avoid bullet points,
    bolding, numbered lists, and Markdown formatting.

    \item Only include the reply in your msg, no other text.
\end{itemize}

\smallskip

The software engineer has a lot of programming background and always
includes technical terms and jargon in their conversation.

\end{tcolorbox}


\par\vspace{\promptgap}

\textbf{(Fc) Designer}\par
\vspace{\promptlabelgap}

\begin{tcolorbox}[
    enhanced,
    width=\linewidth,
    colback=gray!10,
    colframe=gray!50,
    fontupper=\footnotesize\ttfamily,
    boxrule=0.5pt,
    arc=2mm,
    top=2mm,
    bottom=2mm,
    left=2.5mm,
    right=2.5mm
]

Imagine you are working on a data dashboard for a job platform
(e.g., similar to LinkedIn or Handshake analytics).

\smallskip

Users (e.g., recruiters or hiring managers) can browse a live feed
of candidates or applications coming in. New data is continuously
added as candidates apply, update profiles, or interact with the
platform.

\smallskip

The designer wants to introduce a new feature:

\smallskip

A dynamic filtering interface where users can filter candidates by
attributes (e.g., skills, experience level, location), and see results
update immediately as they adjust filters, even as new data is coming in.

\smallskip

For example:

\begin{itemize}[
    leftmargin=1.5em,
    labelsep=0.5em,
    itemsep=0.15em,
    topsep=0.25em,
    parsep=0pt,
    partopsep=0pt
]
    \item A recruiter applies filters such as skill (e.g., Python)
    and years of professional experience (e.g., 3+ years)

    \item As new candidates apply, they automatically appear in the
    filtered list

    \item When the recruiter adjusts filters, results update instantly
\end{itemize}

\smallskip

As a software engineer, your goal is to:

\begin{itemize}[
    leftmargin=1.5em,
    labelsep=0.5em,
    itemsep=0.15em,
    topsep=0.25em,
    parsep=0pt,
    partopsep=0pt
]
    \item Ensure the system is stable and scalable

    \item Avoid overloading servers or slowing down performance

    \item Keep the system efficient as the number of users and data grows
\end{itemize}

\end{tcolorbox}


\clearpage

%
%


\centering
\textbf{(Fd) Software Engineer}\par
\vspace{\promptlabelgap}

\begin{tcolorbox}[
    enhanced,
    width=\linewidth,
    colback=gray!10,
    colframe=gray!50,
    fontupper=\footnotesize\ttfamily,
    boxrule=0.5pt,
    arc=2mm,
    top=2mm,
    bottom=2mm,
    left=2.5mm,
    right=2.5mm
]

You are talking about web development with an experienced software
engineer.

\smallskip

The engineer will walk you through a web development feature they are
adding.

\smallskip

Your goal is to:

\begin{itemize}[
    leftmargin=1.5em,
    labelsep=0.5em,
    itemsep=0.15em,
    topsep=0.25em,
    parsep=0pt,
    partopsep=0pt
]
    \item Ask clarifying questions when something is unclear

    \item Explain concepts back in your own words to confirm your
    understanding

    \item Explore tradeoffs, edge cases, and real-world applications
    by asking follow-up questions
\end{itemize}

\end{tcolorbox}


\par\vspace{\promptgap}

\textbf{(Fe) Designer}\par
\vspace{\promptlabelgap}

\begin{tcolorbox}[
    enhanced,
    width=\linewidth,
    colback=gray!10,
    colframe=gray!50,
    fontupper=\footnotesize\ttfamily,
    boxrule=0.5pt,
    arc=2mm,
    top=2mm,
    bottom=2mm,
    left=2.5mm,
    right=2.5mm
]

I want to design a dynamic filtering interface with real-time updates.
Is it buildable? What could be potential technical issues in building
this? And how long will it take?

\end{tcolorbox}


\newpage
\textbf{(Ff) Software Engineer}
\begin{tcolorbox}[
    enhanced,
    width=\linewidth,
    colback=gray!10,
    colframe=gray!50,
    fontupper=\footnotesize\ttfamily,
    boxrule=0.5pt,
    arc=2mm,
    top=2mm,
    bottom=2mm,
    left=2.5mm,
    right=2.5mm
]

Hey, so I've been working on the real-time filtering feature for the
recruiter dashboard. Here's where I'm at technically:

\smallskip

On the backend, I set up a WebSocket connection using Socket.IO so the
server can push new candidate entries to the client the moment they hit
the database, without the recruiter needing to refresh. For the filtering
logic, I'm running queries against a PostgreSQL database with indexed
columns on skills, experience level, and location \textemdash{} so filter
updates trigger a new parameterized query and stream the results back
through the socket.

\smallskip

On the frontend, I have a React component that holds the active filters
in state. Every time a filter changes, it emits a filter\_update event to
the server, which re-runs the query and pushes back the filtered results.
New candidates that come in while filters are active get checked
server-side against the current filter state before being pushed
\textemdash{} so recruiters only see relevant entries in real time.

\smallskip

One thing I haven't figured out yet is how to handle the UI when a burst
of new candidates comes in \textemdash{} right now it just appends them
to the top of the list, which could be jarring. I also haven't thought
through how the filter panel itself should be laid out or how much of
this should be visible to the recruiter at once.

\smallskip

What are your thoughts on the design side?

\end{tcolorbox}


\end{document}